\documentclass[%
 aip,
 amsmath,amssymb,
 reprint,%
]{revtex4-2}

\usepackage{graphicx}
\usepackage{dcolumn}
\usepackage{bm}
\usepackage[section]{placeins}  

\usepackage{xr}
\usepackage[utf8]{inputenc}
\usepackage[T1]{fontenc}
\usepackage{mathptmx}
\usepackage{etoolbox}
\usepackage{subcaption}
\usepackage{xcolor, soul}
\usepackage[normalem]{ulem}

\newcommand{\changed}[1]{#1}
\renewcommand{\sout}[1]{}

\DeclareUnicodeCharacter{2062}{ }
\makeatletter
\def\@email#1#2{%
 \endgroup
 \patchcmd{\titleblock@produce}
  {\frontmatter@RRAPformat}
  {\frontmatter@RRAPformat{\produce@RRAP{*#1\href{mailto:#2}{#2}}}\frontmatter@RRAPformat}
  {}{}
}%
\makeatother
\begin{document}


\title[]{Explicit core-hole single-particle methods
for $L$- and $M$-edge X-ray absorption and electron energy-loss spectra}
\author{Esther A. B. Johnsen}
\email{esther.johnsen@livmats.uni-freiburg.de}
\affiliation{ Cluster of Excellence livMatS @ FIT, Georges-Köhler-Allee 105, 79110 Freiburg, Germany}%
\affiliation{ 
FIT–Freiburg Center for Interactive Materials and Bioinspired Technologies, University of Freiburg, Georges-Köhler-Allee 105, 79110 Freiburg, Germany}%

\author{Naoki Horiuchi}
\affiliation{ Cluster of Excellence livMatS @ FIT, Georges-Köhler-Allee 105, 79110 Freiburg, Germany}%
\affiliation{ 
FIT–Freiburg Center for Interactive Materials and Bioinspired Technologies, University of Freiburg, Georges-Köhler-Allee 105, 79110 Freiburg, Germany}%

\author{Toma Susi}
\affiliation{University of Vienna, Faculty of Physics, Boltzmanngasse 5, 1090 Vienna, Austria}

\author{Michael Walter}%
\affiliation{ Cluster of Excellence livMatS @ FIT, Georges-Köhler-Allee 105, 79110 Freiburg, Germany}%
\affiliation{ 
FIT–Freiburg Center for Interactive Materials and Bioinspired Technologies, University of Freiburg, Georges-Köhler-Allee 105, 79110 Freiburg, Germany}%
\affiliation{Fraunhofer IWM, MikroTribologie Centrum $\mu$TC, 79110 Freiburg, Germany.}

\date{\today}

\begin{abstract}
Single-particle methods based on Kohn-Sham unoccupied states 
to describe near-edge X-ray absorption (XAS) spectra are routinely
applied for the description of $K$-edge spectra, as there is no complication
due to spin-orbit (SO) coupling.
$L$- and $M$-edge spectra are often addressed via
variants of time-dependent density functional theory (TDDFT) 
based on SO calculations.
Here, we present a computationally efficient implementation based on 
single-particle calculations with core holes within
the frozen-core approximation.
Combined with a semiempirical energy shift and a fixed 
spin-orbit splitting for each core level, this allows for a computationally cheap, while overall accurate
prediction of experimental
spectra on the absolute energy scale.
The spectra are compared to about 40 times slower linear-response TDDFT calculations for molecules
and show similar or even better match with experiment.
An exception are multiplet effects that 
we analyze in detail and show that they cannot be covered by 
a single-particle approximation.
A similar picture emerges for solids, where good qualitative and sometimes even quantitative
agreement to experimental XAS and electron energy-loss spectra 
is achieved.
\end{abstract}

\maketitle

\section{Introduction}

X-ray absorption (XAS) is a widely used method to evaluate 
characteristics of materials where the increasing availability of experimental X-ray sources of high brilliance boosts its use\cite{muller_sulfur_2014,zeng_ultrahigh_2023,matsidik_effect_2023}.
There are different regions within an XAS spectrum that are conveniently viewed 
by different interpretations. 
The region at the onset of absorption, i.e. from the edge up to a few eV,
represents the near-edge X-ray fine structure (NEXAFS) or X-ray absorption near-edge structure (XANES).
This part of the spectrum, as mainly addressed in our work, is sensitive to local chemistry as it probes low-lying conduction-band states in solids or low-lying empty orbitals in molecules.

Higher-energy transitions are known as the extended X-ray absorption ﬁne structure (EXAFS), which is often interpreted in a scattering picture\cite{rehr_theoretical_2000}, giving structural information through Fourier transforms.
This renders EXAFS and XANES useful for structure determination even when lifetime broadening forbids observation of detailed near-edge structures\cite{zeng_ultrahigh_2023}.

In contrast to X-ray photoelectron spectroscopy (XPS) that is
very surface sensitive due to the short path length of the
escaping photoelectron (few\cite{stevie_introduction_2020} nm),
XAS has a penetration depth of several tens\cite{peth_near-edge_2008} of nm.

Interestingly, electron energy-loss spectroscopy (EELS) signals are very similar to 
XAS for small momentum transfers typical for on-axis detectors with a small acceptance angle\cite{Egerton_2008,L_ffler_2011}.
Whereas recording high-resolution XAS spectra usually requires synchrotron radiation, 
EEL spectra can be conveniently measured in
scanning transmission electron microscopes (STEMs), giving rise to the idea of a
synchrotron in a microscope\cite{brown1997synchrotron, Egerton_2008}. Especially with
modern electron monochromators\cite{krivanek2014vibrational}, STEM-EELS has become an increasingly powerful materials characterization tool, with a spatial resolution down to single atoms\cite{susi2017single-atom}.

While in the photoelectron peaks in XPS are assigned to
the core states they originate from,
XAS (and EELS) data is traditionally labeled as $K$-, $L$-, and $M$-edges,
corresponding to core orbitals with respective
principal quantum numbers $n=1,2,3$.
These edges represent sudden increases in the energy-dependent absorption cross-section 
corresponding to a resonance due to the
respective core-state\cite{rehr_theoretical_2000,de_groot_2p_2021}. 

NEXAFS as well as the EELS near-edge fine structure 
can be interpreted within the single-particle picture as 
the excitation of a core electron to an empty state.
As there are many empty states available, the 
NEXAFS spectrum originating from a single core 
orbital usually consists of several peaks,
such that theoretical and computational support is indispensable
for interpretation\cite{kock_nexafs_2023}. 

There are several possible approaches for the simulation of 
XAS spectra as summarized in comprehensive reviews from different viewpoints\cite{besley_density_2020,besley_modeling_2021,norman_simulating_2018,klein_nuts_2021,rehr_theoretical_2000}.
The most accurate calculations are based on equation of motion-coupled cluster theory
(EOM-CCSD), providing reliable spectra both in terms of shape\cite{CC_2008,Vidal2020}
as well as excitation energies\cite{coe_multireference_2015,fransson_xaboom_2021}
despite of problems with orbital relaxation errors\cite{comeau_equation--motion_1993,simons_transition-potential_2021}.
However, such approaches are only applicable to smaller molecules due to their extreme
computational cost and the resulting scalability limitations.

The workhorse for describing 
excitations in larger systems is density functional theory (DFT).
Here, the involvement of (highly) excited states suggests
the use of time-dependent DFT (TDDFT)\cite{kasper_ab_2020,konecny_accurate_2022}.
However, TDDFT is also computationally expensive, and it has limitations
in particular for the description of solids in combination with local- or semi-local functionals\cite{Byun_2020}.
An alternative to TDDFT are single-particle core-hole methods\cite{klein_nuts_2021,prendergast_x-ray_2006,triguero_separate_1999} where the unoccupied
states in the field of a core hole are interpreted as excitations.
These methods are usually applied to $K$-edge spectra\cite{triguero_separate_1999,ljungberg_implementation_2011,michelitsch_efficient_2019,kock_nexafs_2023,klein_nuts_2021,michelitsch_efficient_2019}
due to the difficulties of dealing with the angular momentum of the
initial state and with spin-orbit (SO) coupling\cite{klein_nuts_2021}, although several
$L$-edge EELS simulations using the single-particle density of states\cite{morris2012optados} have notably been reported\cite{nicholls2013probing,Ramasse_2013,hardcastle2017robust,susi2017single-atom}.

In solid systems, excited states that go beyond the single-particle picture
can be described via configuration interaction methods\cite{ikeno_multiplet_2009}
or by solving the Bethe-Salpeter equation\cite{laskowski_understanding_2010,SrTiO3_xas_2023}.

In this article, we apply single-particle methods to simulate $L$- and $M$-edge NEXAFS spectra, specifically focusing on 2$p$ and 3$d$ transitions. 
We compare our approach to experiment for molecules as well as for solids in order to
explore their strengths and weaknesses, and also compare to TDDFT spectra in the case of molecules.
The manuscript first discusses the picture underlying the description of NEXAFS spectra with an emphasis
on comparing the transition from the many-particle to the single-particle viewpoint.
A simple, but effective treatment of SO coupling as a constant shift is also presented.
Then the predictions of the methods are compared to experiment both for molecules as well as
for solid systems. After showing that our methodology works extremely well also for single-atom EEL spectra, we finish with conclusions.

\section{X-ray absorption}

\begin{figure}[!h]
    \includegraphics[width=\linewidth]{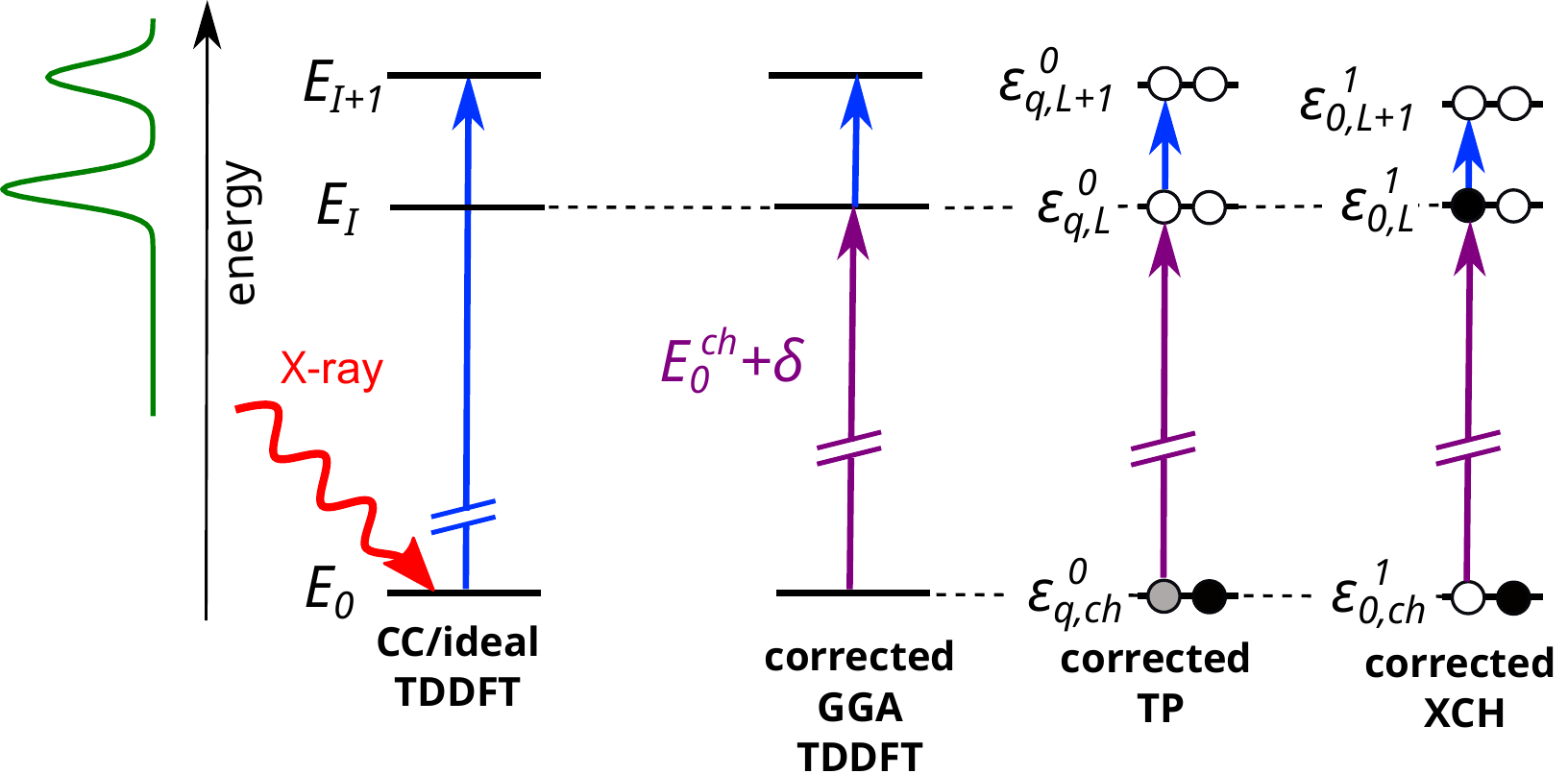}
    \caption{
    \label{fig:energies} Coupled cluster (CC)/ideal TDDFT energetics compared to strategies to obtain X-ray absorption spectra applied in this work.
    Many-particle energies are denoted as $E_0$ for the electronic ground states and $E_{I, I+1}$ for the 
    first and second core-excited state, respectively.
    $E_0^{ch}$ is the energy of the electronic "ground state" in presence of a core-hole. 
    Single particle energies as denoted as $\varepsilon^Q_{q,i}$, where $Q$ is the occupation of the lowest unoccupied molecular orbital ($i=L$) and $q$ the occupation of the core state ($i=\mathrm{ch}$).
    The alignment of the first peak energy to the corrected $\Delta$-Kohn Sham energy (see text) $E^{\rm ch}_0+\delta$ 
    is indicated by the horizontal dashed lines.
    }
\end{figure}
The process of X-ray absorption can be interpreted in different approximations,
as depicted schematically in Fig.~\ref{fig:energies}.
Generally, an X-ray photon is completely absorbed by the molecule or solid under observation,
leading to a highly excited state of the system due to the large photon energy.

We first discuss the process within the most accurate many-particle picture
in terms of many particle ground and excited states.
The absorption cross-sections are proportional to the
unitless oscillator strength\cite{prendergast_x-ray_2006,walter_time-dependent_2008,hilborn_einstein_1982,walter_photoelectron_2008}
\begin{equation}
    (f_{I})_\eta = \frac{2m_e}{\hbar^2e^2} (E_I-E_0)\left|\left\langle\Psi_I\right|\hat{\mathbf{u}}_\eta\cdot\mathbf{D}\left|\Psi_0\right\rangle\right|^2,
    \label{eq:ffi_manyparticle}
\end{equation}
where $m_e$ is the electron mass, $e$ the unit charge, $\hbar$ the reduced Planck constant and 
$\eta=x,y,z$ are Cartesian coordinates. 
$E_0$ is the energy of the ground state, $E_I$ the energy of the core-excited state
of index $I$ (there are many excited states forming the full XAS-spectrum)
and $\Psi_0, \Psi_I$ denote their corresponding
many-electron wave functions. 
The 
dipole operator in length form $\mathbf{D}$ contains the contributions of all electrons. 
The relevant coordinate $\eta$ is chosen by the polarization of the X-ray 
photon $\hat{\mathbf{u}}_\eta$.
A specific $\hat{\mathbf{u}}$ appears in
angle-dependent X-ray spectroscopy\cite{george_long-range_2014} or 
X-ray circular dichroism\cite{norman_simulating_2018}.
A similar expression involving many-electron wave functions 
appears in the GW approximation\cite{gilmore_efficient_2015} resulting in
Eq. (\ref{eq:ffi_manyparticle}) for vanishing
momentum transfer by the absorbed photon.
The target molecules or solids are usually in random orientations,
such that the averaged 
intensity, i.e. $f_{I}=[(f_{I})_x + (f_{I})_y + (f_{I})_z]/3$ is
measured.

The calculation of transition energies and oscillator strengths 
in Eq.~(\ref{eq:ffi_manyparticle})
requires many body energies and wave functions.
Many-electron methods are very demanding
requiring expansion in a large number of 
single particle orbitals or many particle-hole pairs.
Most exact, but also most demanding is the use of
many-body wave functions as provided by e.g.
coupled cluster (CC) descriptions for ground and excited states. 
The equivalent could be achieved by TDDFT in case the exact 
(frequency dependent\cite{cave_dressed_2004}) 
exchange-correlation kernel would be known.
Practical TDDFT calculations, in particular based on gradient-corrected kernels as applied here, do not, however, necessarily lead
to accurate transition energies\cite{m.e._casida_progress_2012}.

A huge simplification is achieved when
XAS is interpreted within the single-particle picture 
as transitions of a core electron to an empty orbital 
(or the conduction band for solids).
The oscillator strength reduces in this picture
to\cite{leetmaa_theoretical_2010}
\begin{equation}
    (f_{fc})_\eta = \frac{2m_e}{\hbar^2e^2} \varepsilon_{fc}\left|\left\langle\phi_f\right|\hat{\mathbf{u}}_\eta\cdot\mathbf{D}\left|\phi_c\right\rangle\right|^2,
    \label{eq:ffi}
\end{equation}
where the matrix element is calculated from single-particle orbitals
of the core state $\phi_c$ and the final state $\phi_f$, and 
the dipole operator simplifies to $\mathbf{D}=-e\mathbf{r}$
with the electron spatial coordinate $\mathbf{r}$.
The transition energies are now Kohn-Sham energy differences
\begin{equation}
\varepsilon_{fc}=\varepsilon_{q, i}^Q - \varepsilon_{q, {\rm ch}}^Q \; .
\label{eq:eps_fc}
\end{equation}
Eq. (\ref{eq:ffi}) is a harsh approximation as the energies and wave-functions
of many-electron excitations are replaced by the energy and orbital of a single, 
seemingly independent unoccupied state, 
that has to describe all excitations to 
itself irrespective where the excitation came from.
We will see below, that this approximation indeed breaks down for strongly correlated 
systems\cite{quintanilla_strong-correlations_2009}.

These single-particle energies
may be again evaluated in different approximations that 
mainly differ by 
the occupation of the core state $q$ as well as
by the occupation $Q$ of the lowest unoccupied state (LUMO).
These choices not only influence the energy values, but also change the single-particle states
in Eq.~(\ref{eq:ffi}), where we have suppressed the labels $q,Q$ for clarity.
More details about the implementation can be found in the Supplementary Material
(SM) Sec. II.

Regardless of whether single-particle or many-particle pictures are adopted,
the calculated spectra are usually rigidly shifted
to match their lowest energy peak with that of the experiment\cite{Kasper2018}.
With the exception of some reported computationally expensive EOM-CCSD calculations\cite{kasper_ab_2020}
these shifts often amount to several eV.

Similar shifts appear in gas-phase XPS, where some of us have developed a semi-empirical method
to obtain these from experimental data\cite{walter_offset-corrected_2016}.
In this method, we calculate the energy
of the lowest energy transition
as the energy difference between
two DFT ground state calculations. 
One is the ground state energy $E_0$ and the other is the energy of the
electronic ground state in the presence of the core-hole $E_{ch}^0$.
The zero indicates the overall electrical neutrality of this configuration,
i.e. it contains an extra valence electron as relevant for XAS.

Despite the principal difficulty of describing such a high-energy core-excited state within 
DFT ground-state theory to obtain $E_{ch}^0$, the similarity 
to the equivalent-core approximation\cite{norman_simulating_2018} 
rationalizes this approach.
The first transition energy, known as $\Delta$-Kohn Sham energy ($\Delta$KS), is then
\begin{equation}
    E_B^0 = E^0_{\rm ch} - E_0 \; .
    \label{eq:EB0}
\end{equation}
Local and semilocal functionals are rather accurate in the evaluation of ionization potentials
and electron affinities from total energy differences\cite{dabo_koopmans_2010},
even better than computationally much more demanding 
GW approximations\cite{rostgaard_fully_2010}.
Nevertheless, the core-hole energies calculated from Eq.~(\ref{eq:EB0}) 
are observed to be shifted against the experiment, where the shift depends
on the core state and the functional, but not on the chemical 
environment\cite{walter_offset-corrected_2016,kock_nexafs_2023}.
This makes it possible to obtain a semi-empirical correction by comparing 
a large body of calculations with experimental data $E_B^{\rm exp}$ 
to obtain shift values 
\begin{equation}
    \delta = E_B^{\rm calc} - E_B^{\rm exp}
    \label{eq:delta}
\end{equation}
that can be used to correct the raw 
calculated energies $E_B^{\rm calc}$ as indicated in Fig.~\ref{fig:energies}.
This method has successfully predicted the XPS spectra of 
ionic clusters \cite{walter_experimental_2019} and carbon materials\cite{walter2019fermilevelpinningdefects,walter_origin_2025}, 
as well as to S 1$s$ NEXAFS spectra \cite{kock_nexafs_2023,matsidik_effect_2023}.
We will see below that there is also substantial uncertainty in experimental
absolute energies. The shift $\delta$ can therefore be seen as an
average over experimental knowledge.
The shift depends on core state and depends on the 
exchange-correlation functional. Optimizing the shift to experiment 
therefore also corrects for inaccuracies of the functional.
There are hints that at least a large part of the shift stems from
the use of the frozen-core approximation, as all-electron calculations
seem to have negligible shifts at least for light elements\cite{susi_calculation_2015,kahk_accurate_2019}.

In contrast to $K$ edges ($s$-type core states), $L$ and $M$ edges
have finite orbital angular momenta and are thus
subject to spin-orbit (SO) coupling\cite{jacob_spin_2012}.
The SO coupling arises from the relativistic Dirac equation
and couples the spin quantum $s$ to the orbital
angular momentum $l$, leading to two different core-state energies
for the same orbital.
This SO splitting is larger for orbitals of high binding energy,
while it has comparably small effects on the 
valence or empty (conduction band) states.
The SO splitting usually also depends negligibly on the chemical 
environment\cite{Vidal2020} and $\delta$ from Eq.~(\ref{eq:delta}) 
applies to both SO split energies.
This is equivalent to defining one shift $\delta$ for each
SO split state, where the two $\delta$ are separated by 
the SO splitting.
The contributions of the two
states are then approximated as the same spectrum separated in energy by
the SO splitting and weighted according to the relative
spin multiplicities\cite{manyakin_electronic_2019,walter_experimental_2019}.

Having obtained an accurate value for the first transition energy, i.e. to the
LUMO in NEXAFS spectra, the question appears how to describe higher-lying states (c.f. Fig.~\ref{fig:energies}).
The natural choice in the framework of DFT seems to be TDDFT, as often applied in the literature but
with mixed success. 
TDDFT is usually based on ground-state orbitals, where it can only describe single excitations if
the usual energy-independent kernels are used\cite{cave_dressed_2004}. 
The orbitals in the field of the core-hole may be better suited to describe the final state
after photon absorption\cite{norman_simulating_2018}. 
These orbitals
already represent multiple excitations with respect to the ground-state orbitals
and have been demonstrated to improve TDDFT XAS calculations\cite{carter-fenk_electron-affinity_2022}.
Along these lines, methods that consider orbital changes under excitation\cite{ivanov_method_2021,hait_orbital_2021}
 are very successful and sometimes even exceed standard TDDFT accuracy
in practical calculations.

$K$-edge NEXAFS spectra have therefore been successfully calculated within a single-particle 
picture, where the corresponding orbitals are obtained within the field of an explicit
core-hole. 
There are different approximations possible\cite{klein_nuts_2021}, and it has turned out that the transition-potential (TP) method\cite{triguero_separate_1999,ljungberg_implementation_2011,susi_core_2014}
with $Q=0$ and $q=1/2$ in Eq.~(\ref{eq:eps_fc}) often
leads to a rather good agreement with experiment\cite{leetmaa_theoretical_2010,kock_nexafs_2023}.
There are rationalizations for other
$q$ values\cite{Hirao_2021,hirao_core_2023} in extension of Slater’s original transition state
concept. The latter is based on setting an occupation of $1-q$ 
in the excited orbital of interest, which is further approximated in the TP method to $Q=0$ in order 
to create a single potential for all excitations\cite{triguero_separate_1999}.
In the viewpoint of the various approximations adopted, we generally use $q=0.5$ as in original TP, but treat it as free parameter in some calculations.
The excited core-hole (XCH) method, where $Q=1$ and $q=0$ in Eq.~(\ref{eq:eps_fc})
as appropriate for a neutral excitation,
was also successfully
applied\cite{prendergast_x-ray_2006,pascal_x-ray_2014}.
The system's neutrality in XCH greatly benefits periodic calculations. 

\section{Computational settings}

The electronic structure of the molecules and solids considered
is described by DFT within the projector augmented wave method\cite{blochl_projector_1994}
as implemented in GPAW\cite{mortensen_real-space_2005,enkovaara_electronic_2010,mortensen_gpaw_2024}. 
The exchange-correlation energy was evaluated as devised by Perdew, Burke and Ernzerhof (PBE)\cite{perdew_generalized_1996}. 
PBE is a robust functional derived from basic principles which is 
applied for molecular systems as well as solids.
We have good experience using PBE for the description of core-hole spectra on the absolute 
energy scale by applying a semi-empirical energy correction fitted for this functional\cite{walter_offset-corrected_2016,walter_experimental_2019,walter_origin_2025,kock_nexafs_2023}.

Molecules were simulated in Dirichlet (non-periodic) boundary conditions, where the simulation cell contained at least 5~\AA\ around each atom.
Solids and the two-dimensional (2D) structures are calculated in periodic boundary conditions.
The 2D systems included at least 10~\AA\ of distance between the layers to avoid their interaction.
All structures were relaxed until all forces were below 0.05~eV/\AA.

GPAW offers the possibility to represent the
smooth part of the Kohn-Sham orbitals on real-space grids, momentum space (plane wave)
grids or linear combination of atomic orbitals.
The electron density is represented on real-space grids, where we chose a
grid spacing of 0.1~\AA\ in our calculations.
The Kohn-Sham orbitals are also represented on real-space grids with 0.2~\AA\ grid-spacing.
The Kohn-Sham orbitals in the 2D system were represented by plane waves with
an energy cut-off of 600~eV.
The use of grids in real or momentum space (plane waves) offers a facile and smooth 
convergence to the complete basis set limit\cite{mortensen_real-space_2005,wurdemann_density_2015}.
Equidistant grids also allow for very effective parallelization.\cite{enkovaara_electronic_2010}

We use the frozen-core approximation\changed{\cite{mortensen_real-space_2005,enkovaara_electronic_2010}} for orbitals not participating to
the valence band, which are calculated for free atoms.
This approximation allows to use coarser grids
while still catching all orbitals necessary to cover chemical interactions. In explicit the frozen core states are 1s$^2$=[He] for C and O, [He]2s$^2$2p$^6$=[Ne] for Si, S, Cl and Ti, 
[Ne]3s$^2$ for Ni, [Ne]3s$^2$3p$^6$3d$^{10}$=[Ar]3d$^{10}$ for Sr, and [Ar]3d$^{10}$4s$^2$4p$^6$=[Kr] for Sn. 

The frozen-core approximation has the additional benefit in that it allows to describe the presence of a 
core hole obtained by a self-consistent free-atom 
calculation with a reduced occupation $q$ of the 
core state\cite{ljungberg_implementation_2011}.
This approach explicitly considers a localized core hole and
thus avoids problems with delocalized core-hole states 
in symmetric environments\cite{klein_nuts_2021}, 
despite that solutions to this problem exist\cite{yu_accurate_2025,oosterbaan_non-orthogonal_2018}.
The price to pay is the absence of core-hole \changed{state} relaxation 
\changed{in a chemical environment different to an isolated atom}
changing the energetics\cite{susi_calculation_2015}.
This is considered and corrected by our semiempirical shifts that were 
fitted for these fixed core-holes.
Details about the calculation of the shifts $\delta$ [Eq.~(\ref{eq:delta})]
are found in SM Sec.~III.

We contrast these single-particle calculations with linear-response TDDFT (LrTDDFT) 
calculations\cite{walter_time-dependent_2008}, where we include the relevant 
core-states into the valence orbitals. 
There is thus no explicit core-hole present in
our LrTDDFT that is purely based on Kohn-Sham ground-state orbitals.
LrTDDFT in combination with gradient-corrected kernels as applied here 
though the PBE functional is known have severe problems in describing charge-transfer
excitations.\cite{Carter-Fenk2021} We therefore contrast 
selected spectra calculated with a range-separated functional based on the Yukawa potential (LCY-PBE)\cite{Seth2012,wurdemann_charge_2018} which is known to perform better in this respect\cite{wurdemann_charge_2018}. The Yukawa damping parameter is obtained by 
requiring the equality of the highest molecular orbital's energy with the negative of the ionization potential\cite{livshits_well-tempered_2007,Joo2018} as valid in exact DFT\cite{almbladh_exact_1985}.

The SO splittings for each element are calculated through perturbation theory\cite{olsen_designing_2016,mortensen_gpaw_2024}
for free atoms as detailed in SM.
We have checked that the calculated SO-splittings vary very little compared to their values
within the molecules, such that the assumption of a common SO splitting
per element and core-state is a good approximation independent of the chemical environment
(see SM Sec.~IV for details).
The oscillator strengths are calculated without spin-orbit effects and 
the full spectrum is constructed assuming the same spectrum for the two contributions 
separated in energy by the SO splitting and scaled by the corresponding weight as given in Tab. S1 in SM.

The oscillator strengths from Eq.~(\ref{eq:ffi_manyparticle}) calculated via
LrTDDFT or from Eq.~(\ref{eq:ffi}) calculated via DFT use the corrected energies involving
the shifts $\delta$. 
In order to compare the spectra to experiment, these oscillator strengths are folded 
(convolved) by Cauchy distributions (Lorentzians) of variable width modeling to finite lifetime,
vibrational contributions\cite{travnikova_esca_2012} and limited experimental resolution.
The EEL spectra cover a larger range of excitation energies than the XAS spectra and
higher-energy excitations are expected to have shorter lifetimes, resulting
in larger  broadening. This is taken into account by energy dependent broadening
as detailed in SM.
The convolution procedure generally results in 
intensities that are folded oscillator strengths (FOS) with the unit 1/eV.

\section{Results and Discussion}

\subsection{Molecules}

In this section we compare calculated spectra in different approximations to experiment.
We in particular emphasize the comparison of the many-particle spectra determined by LrTDDFT
to fixed core-hole single-particle approximations in the 
XCH (full core hole $q=0$ and extra electron in LUMO $Q=1$)
and the TP (half core-hole $q=1/2$ and $Q=0$) approximations.

\begin{figure}[!h]
    \includegraphics[width=0.51\textwidth]{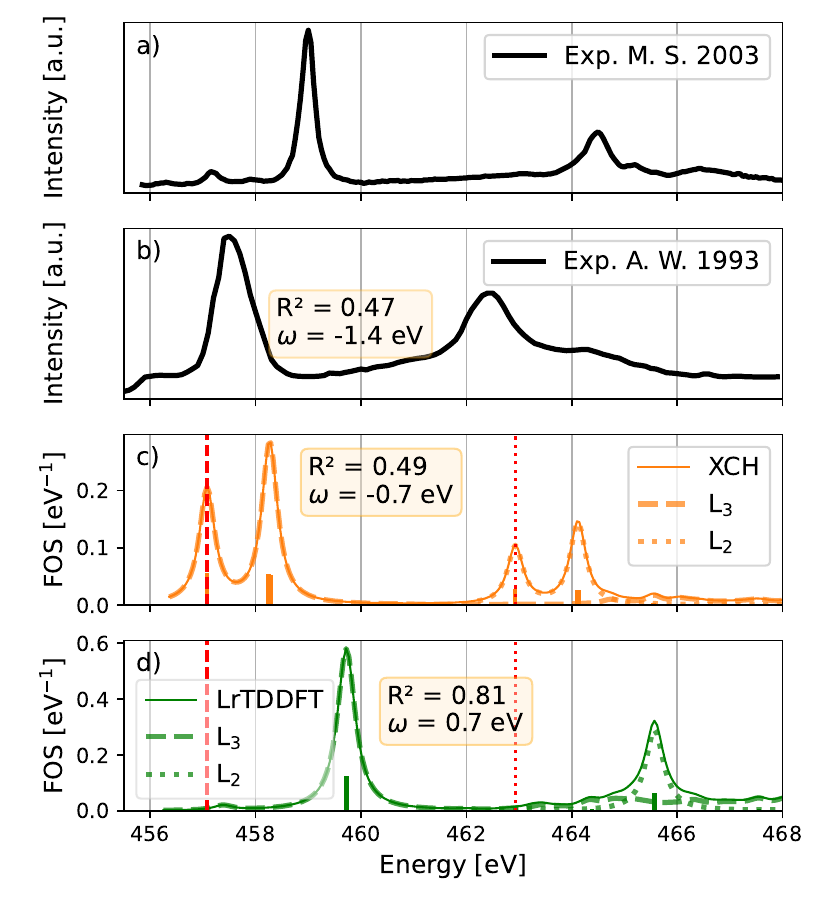}
\caption{Ti 2$p$ or $L_{2,3}$-edge NEXAFS spectra for TiCl$_4$. The experimental spectrum in a) is adapted from Ref.~\onlinecite{Wen_1993} and that in b) from Ref.~\onlinecite{Stener_2003}.
    \changed{The methods used to calculate the spectra in c,d) are indicated in the legends.}
    The oscillator strength of each transition is shown
    as a stick spectrum. The energetic position of the first excitation of the $L_3$ contribution that is aligned to the corrected $\Delta$KS is marked by the dashed vertical red line. The $\Delta$KS of the SO shifted $L_2$ contribution is marked by the dotted vertical red line.
    The folded oscillator strength (FOS)
    is obtained by from the stick spectrum as explained in the text.
    The $L_3$ and $L_2$ contributions are indicated by dashed and dotted lines in c,d), respectively.
    The optimal R$^2$ and the corresponding energy shift $\omega$ applied to the experiments
    is given (the experimental spectra are still depicted
    on their original energy scale, but were shifted to evaluate R$^2$).
    }
    \label{fig: TiCl4}
\end{figure}

The Ti $L_{3,2}$ edge spectra of TiCl$_4$ are presented in 
Fig.~\ref{fig: TiCl4}.
There are two experimental spectra for this molecule available in the literature\cite{Wen_1993, Stener_2003}, which are similar in the appearance of two main peaks, but there
there are significant differences also.
The spectra differ in peak-width (different experimental conditions), absolute peak positions 
(difficulties in determining 
the absolute energy scale), and in the relative energy separation of the peaks.

We generally evaluate the agreement between two spectra by calculating the coefficient of determination R$^2$ value\cite{Spiess_2010,Gleason_2024, Guda_2021}, where a value of unity describes perfect agreement.
We allow for a variable width of the Cauchy distributions
as well as for a variation in absolute energy by a shift 
$\omega$
to optimize agreement with experiment (see SM Sec.~VI for details).
Note that the shift $\omega$ is for a specific experiment, while the shift $\delta$ from Eq.~(\ref{eq:delta})
corrects the calculation to the average of high energy resolution experiments. 
We therefore regard our calculated and corrected
energy as an accurate energy scale to which the specific experiment will be corrected.

We first compare the two experimental spectra for TiCl$_4$ in a similar way with each other. 
Here the more resolved spectrum of 
Stener et al.\cite{Stener_2003} has to be shifted by 
$\omega=-1.4$ eV to maximize $R^2$, still leading to a low $R^2=0.47$.
The energetic distance between the main peaks caused by 
spin-orbit splitting is 5 and 5.5~eV in the two experiments respectively, and therefore also different (our calculated SO splitting value is 5.85~eV for Ti 2$p$).
These disagreements show the experimental difficulties in
the precise determination of energy scales. 

\begin{figure}[!b]
    \includegraphics[width=0.5\textwidth]{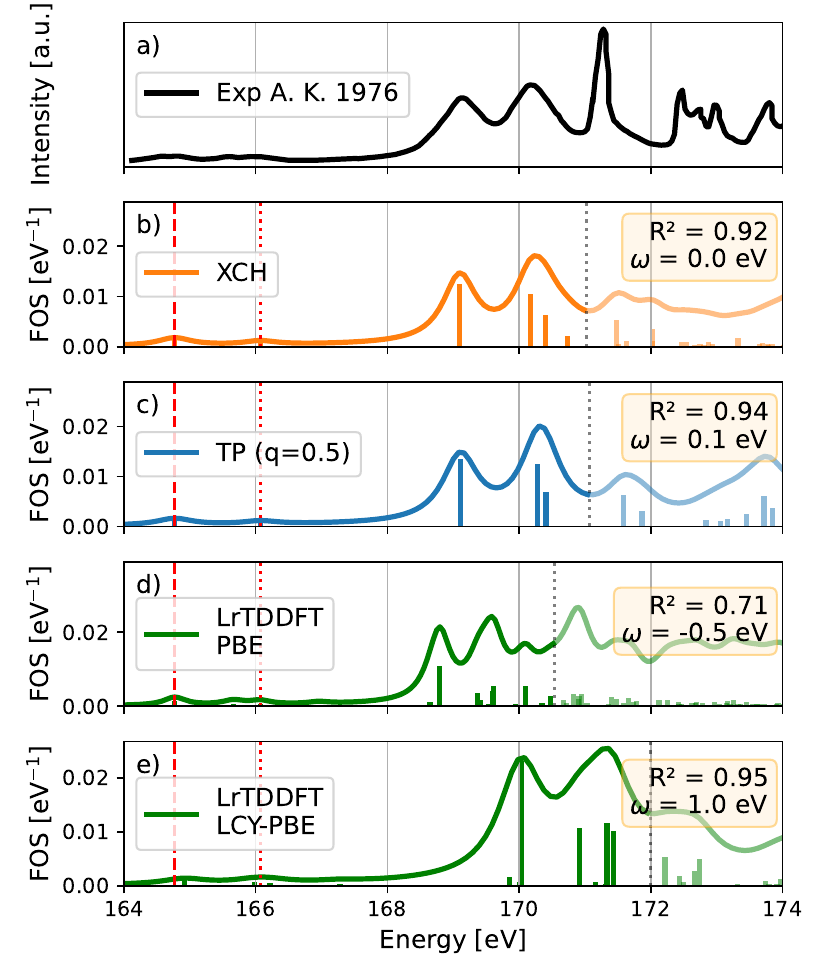}
    \caption{
    \label{fig:SO2} Calculated S 2$p$ or $L_{2,3}$-edge NEXAFS spectra for SO$_2$ and a) an experimental spectrum adapted from Ref. \onlinecite{krasnoperova_fine_1976}.
    \changed{The methods used to calculate the spectra in b-e) are indicated in the legends.}
    The spectral region used for obtaining the R$^2$ value is depicted in solid color separated from the spectral region excluded by the black dotted line (see text for details).
    }
\end{figure}

The better resolved spectrum of Stener et al.\cite{Stener_2003} shows a very small peak, about 2~eV lower than the main feature and well separated $2p^{1/2}$ contributions above 464~eV.
These features are well described by the lrTDDFT calculations 
resulting in a rather large R$^2$.
Similar agreement has been found from related calculations\cite{Stener_2003}.
The energies of the first transitions that are shifted to the corrected $\Delta$KS energy 
for $L_3$
and to $\Delta$KS plus the SO splitting for $L_2$ are
indicated in the figure. This energy notably corresponds to a transition with
very small oscillator strength, which therefore does not appear in the spectrum.
We note, that the shift $\omega=0.7$ eV to align the experimental spectrum with the
LrTDDFT spectrum is smaller than the corresponding alignment between the two experiments.

The single-particle XCH spectrum looks very different, however, and also shows a low R$^2$.
The TP spectrum is very similar and can be found in SM. 
There are two strong peaks in the low-energy region caused by the $2p_{3/2}$ contribution.
These correspond to the unoccupied Ti 3$d$ shell that is split due to the symmetry of the four Cl atoms.
The corresponding 2$p$ to 3$d$ transitions are obviously coupled by LrTDDFT, but this is not the case within the single-particle picture.
This coupling of empty states is known from atomic physics and is dubbed the
multiplet effect in the literature\cite{NiO_xas_GROOT2005}.

\begin{figure*}[htb]
    \includegraphics[width=0.49\textwidth]{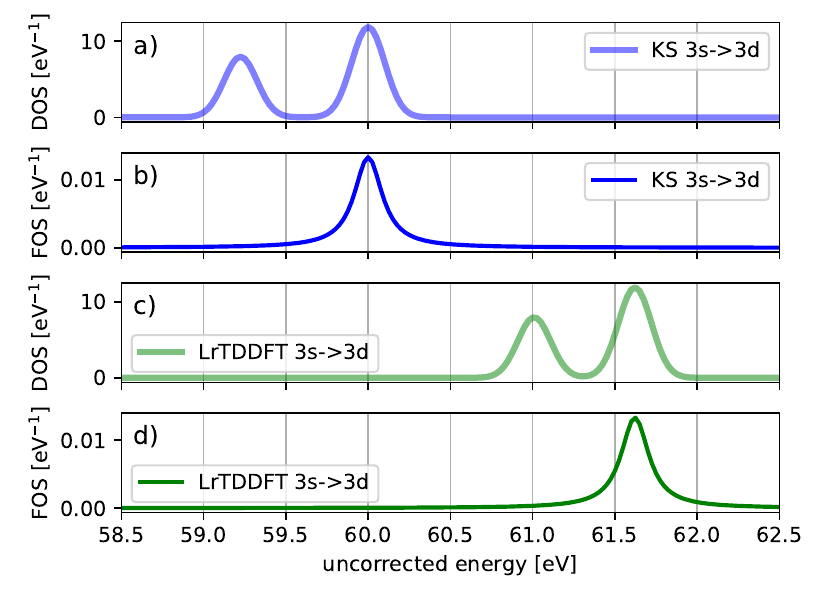}
    \includegraphics[width=0.49\textwidth]{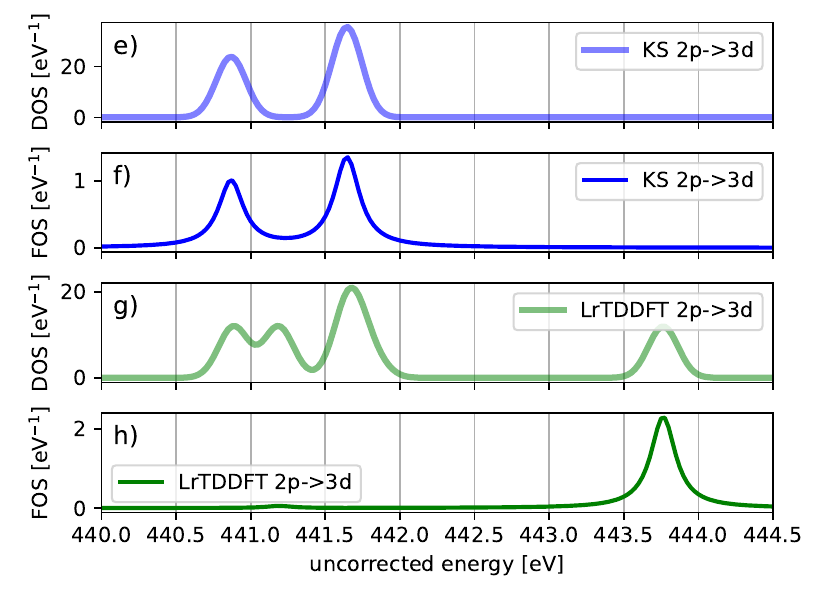}
    \caption{Transitions to the empty 3d states of Ti in TiCl$_4$ as density of states (DOS) and folded oscillator strengths (FOS) folded by Gaussians and Lorentzians of 0.1 eV width, respectively.  a-d) 3s to 3d transitions and e-h) 2p to 3d transitions. KS denotes Kohn-Sham single particle transitions \changed{in a,b,e,f) \sout{and} that are} contrasted to the LrTDDFT transitions \changed{in c,d,g,h)}.
    }
    \label{fig: DOT TDDFT}
\end{figure*}
In order to understand the origin of the multiplet effect, we compare the 2p to 3d transitions to the
3s to 3d transitions in Fig.~\ref{fig: DOT TDDFT}. The density of excited states (DOS) corresponding to 3s to 3d transitions show two peaks due to the splitting of the d-states in the tetrahedral symmetry of the Cl atoms in TiCl$_4$. 
There is some small transition probability for one type of these excitations due to mixing of p-states in this otherwise dipole-forbidden transition. This is reflected both in the single particle transitions (KS) as 
well as in the excitations from LrTDDFT. I.e. TDDFT merely leads to a slight energy shift here.

The situation is remarkably different when comparing KS and LrTDDFT for 2p to 3d transitions in Fig.~\ref{fig: DOT TDDFT} e-h). We restrict to the main contribution 2p$_{3/2}$ contribution, the $2p_{1/2}$ part is found at higher energies. The KS single-electron picture shows two transition energies split like the 3s to 3d transitions as all relevant 2p states have the same energy. 
Now both DOS peaks have non-vanishing oscillator strength in KS giving the two prominent peaks seen in Fig.~\ref{fig: TiCl4}b). The DOS of LrTDDFT is very different in that more peaks are built from the 15
2p to 3d transitions. The most dramatic effect is seen for the oscillator strength that now is practically 
only finite for the highest energy transitions, leading to one strong main peak as seen in Fig.~\ref{fig: TiCl4}a) and in experiment.

We note, that this effect cannot be described within an independent orbital picture. TiCl$_4$ therefore represents a strongly correlated system in the empty $d$ states. 
The excitations in TiCl$_4$ change according to "where they came from", such that there are no independent $d$-orbitals that are able to describe all transitions irrespective from which orbital they are filled.

\begin{figure}[!h]
\centering
\includegraphics[width=0.5\textwidth]{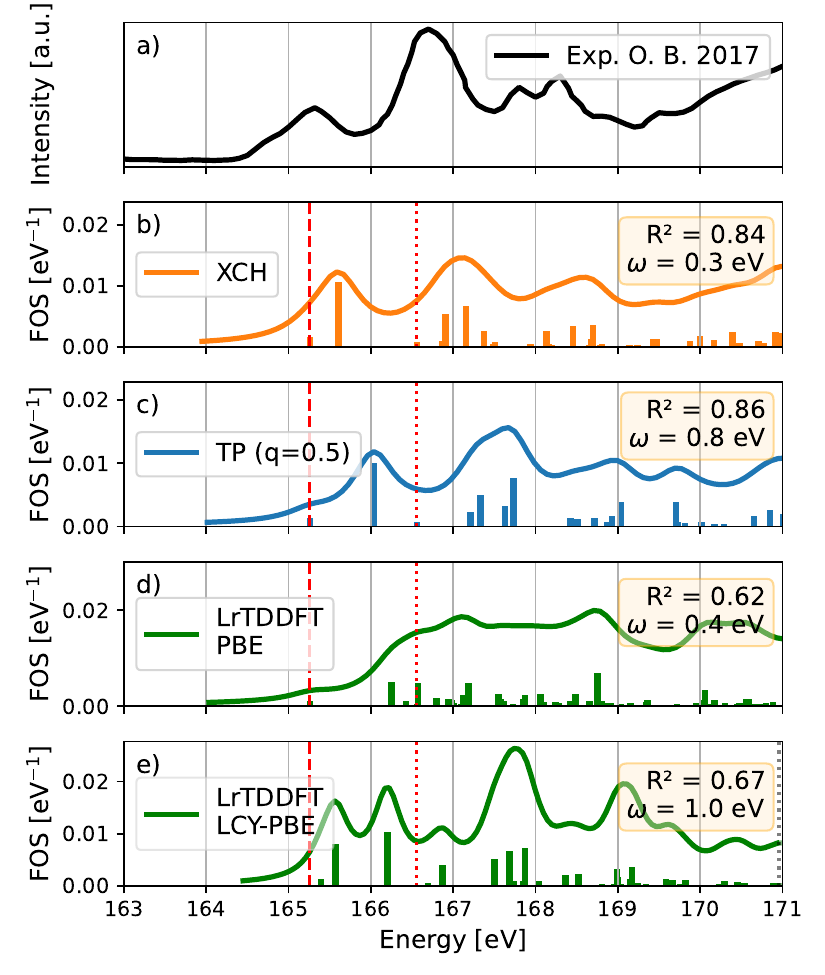}
\caption{\label{fig: Thiophene}
 As Fig.~\ref{fig:SO2}, but for thiophene. The experimental spectrum in a) is adapted from Ref.~\onlinecite{baseggio_s2p_2017}.
 \changed{The methods used to calculate the spectra in b-e) are indicated in the legends.}
}
\end{figure}
The spectrum of the sulfur $L_{2,3}$ edge or S 2$p$ for 
the SO$_2$ molecule is depicted in Fig.~\ref{fig:SO2}.
The experimental spectrum\cite{krasnoperova_fine_1976} shows two shallow peaks on the lower energy side at 164.5 and 166~eV and several peaks with larger intensity starting from 169~eV.
The transitions above 171 eV show a clear peak structure that is assigned to Rydberg states\cite{krasnoperova_fine_1976}.
As we model the vibrational broadening by a fixed folding width, our simulation cannot describe this highly resolved region. 
Vibrationally resolved simulations have been demonstrated\cite{travnikova_esca_2012,stauffert_optical_2019,walter_ab_2020}, but go well beyond our approach here.

All simulated spectra qualitatively describe this general structure of the spectrum, and the positions of the peaks are in overall agreement.
The linear-response TDDFT spectrum is slightly more blurred for the strong transitions than the single-particle approaches.
Similarly to their visual appearance, all methods have a good accuracy, 
but the single-particle approaches are clearly superior to LrTDDFT in terms of the R$^2$ value. 

The broadening in the LrTDDFT spectrum is caused by the gradient-corrected functional PBE we 
are using as is demonstrated by a comparison applying the range separated LCY-PBE functional in Fig.~\ref{fig:SO2}d). The LCY-PBE spectrum is more peaked and at least partly resembles the experimental 
peak structure from experiment.
We note, that the use of Hartree-Fock exchange as required in LCY-PBE is extremely expensive in grid based approaches due to
the need for an explicit calculation of the exchange integrals. These can be calculated partly analytically in basis set calculations, which is not possible here. The LCY-PBE LrTDDFT calculation used 260 CPU days, while the PBE LrTDDFT calculation finished in 9 CPU hours, i.e. three orders of magnitude faster. 
The TP and XCH calculations finished in roughly 14 CPU minutes, giving a speed-up factor of 40 relative to PBE LrTDDFT and 30.000 relative to LCY-PBE and demonstrating the cost-effectiveness of the single-particle methods.

The calculated $L_{2,3}$ edge spectra of thiophene (C$_4$H$_4$S) are compared to the experiment
in Fig. \ref{fig: Thiophene}.
Similar to SO$_2$,
we observe a generally good agreement when using core-hole methods, but the lrTDDFT spectrum is very blurred, such that no peak structure is visible anymore.
This is reflected in its lower value $R^2$ compared to TP and XCH.
The experimental spectrum has to be shifted by 0.3--0.8~eV to maximize $R^2$, i.e., slightly more than
in the case of SO$_2$.

\begin{figure}[b!]
    \includegraphics[width=0.5\textwidth]{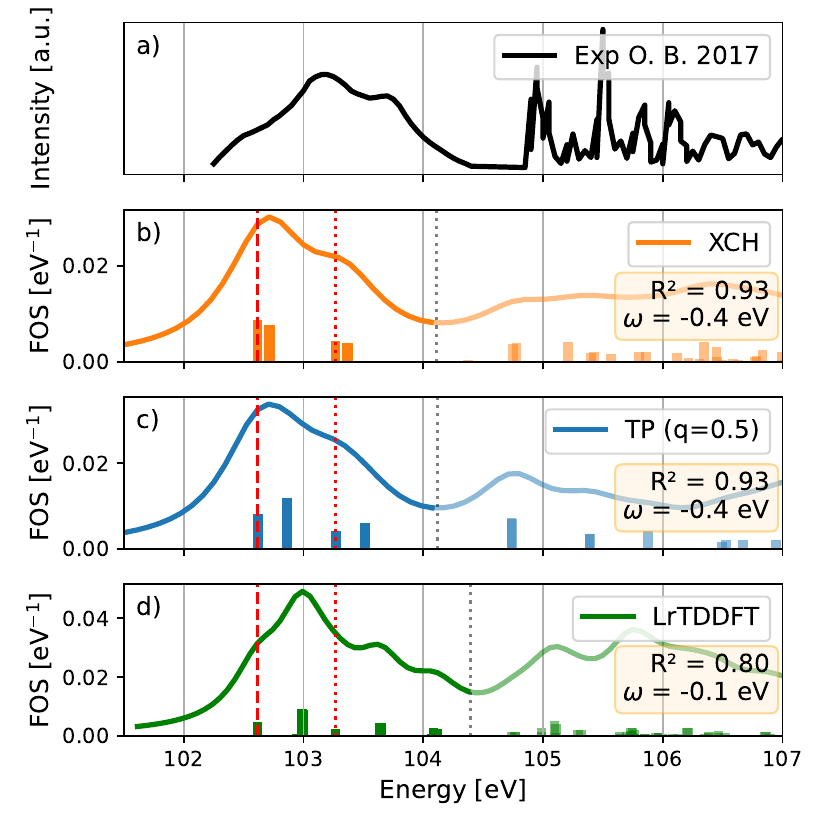}
    \caption{
     Si 2$p$ or $L_{2,3}$-edge NEXAFS spectra for SiH$_4$. 
     The experimental spectrum is adapted from Ref.~\onlinecite{P_ttner_1997}. 
     \changed{The methods used to calculate the spectra in b-d) are indicated in the legends.}
     The spectral region used for obtaining the R$^2$ value and
    the corresponding shift $\omega$ is depicted in solid colour (see text for details).
    }
    \label{fig: xas SiH4}
\end{figure}
Next, we turn to the $L_{2,3}$ edge of silicon, where we first address the spectrum of SiH$_4$
in Fig.~\ref{fig: xas SiH4}. The experimental spectrum can be separated into two regions,
where the transitions below 104.5~eV are largely broadened. 
The transitions above this energy show a clear peak structure that is assigned to the vibrational sidebands of the Rydberg states\cite{P_ttner_1997}.
Similarly to the higher-energy peaks in SO$_2$, the assumption
of a fixed folding hinders the description of these narrow transitions.

The calculations agree rather well to experiment in the first broad peak and also describe the less intense (after averaging in the experiment) higher-energy region. Taking only the lower-energy peak into account leads to rather large $R^2$ values, where the single-particle approaches outperform LrTDDFT.
Only small shifts must be applied to the experimental spectrum to maximize $R^2$.

\begin{figure}[h!]
    \includegraphics[width=0.5\textwidth]{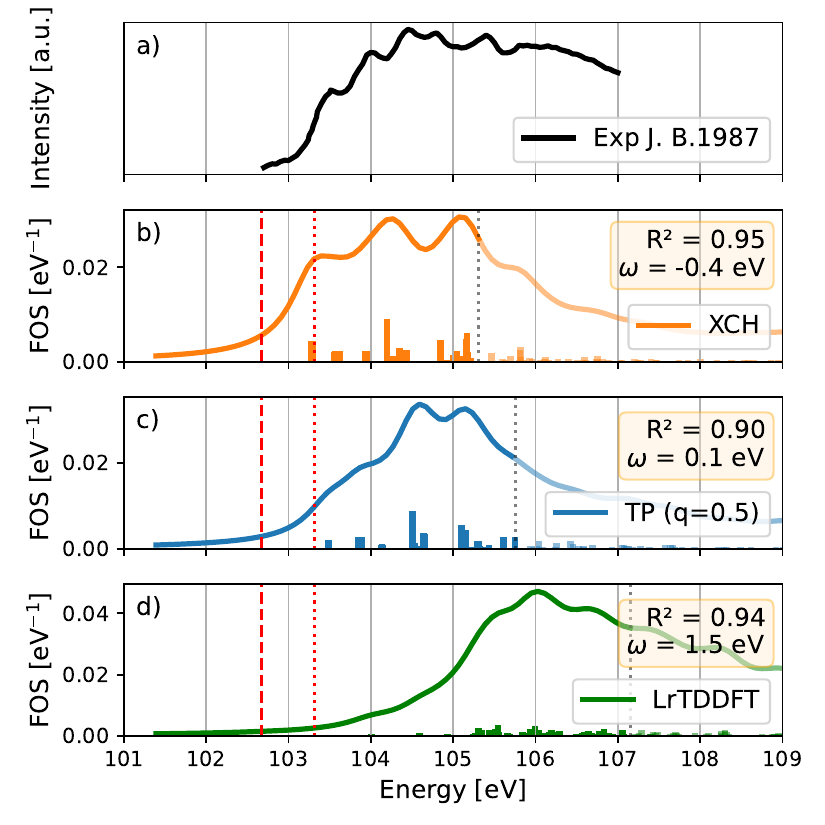}
    \caption{Si 2$p$ or $L_{2,3}$-edge NEXAFS spectra for Si(CH$_3$)$_4$. 
     The experimental spectrum \changed{in a)} is adapted from Ref.~\onlinecite{BOZEK198733}.
     \changed{The methods used to calculate the spectra in b-d) are indicated in the legends.}
     }
    \label{fig: xas SiMe4 lrtddft}
\end{figure}
The $L_{2,3}$ edge spectra of Si(CH$_3$)$_4$ are compared in Fig.~\ref{fig: xas SiMe4 lrtddft}.
The experimental spectrum mainly consists of a broad peak with some visible substructure.
This is reflected in all simulations, but the shifts of this broad peak relative
to the first transition at $\Delta$KS are quite different. 
LrTDDFT shows the largest shift of
this broad peak to higher energies. 
This is reflected in the largest shift $\omega$ needed to maximize the agreement of LrTDDT compared to TP and XCH. 
The flexibility of these different shifts results in similar $R^2$ values for all models.

\begin{figure}[!h]
    \includegraphics[width=0.5\textwidth]{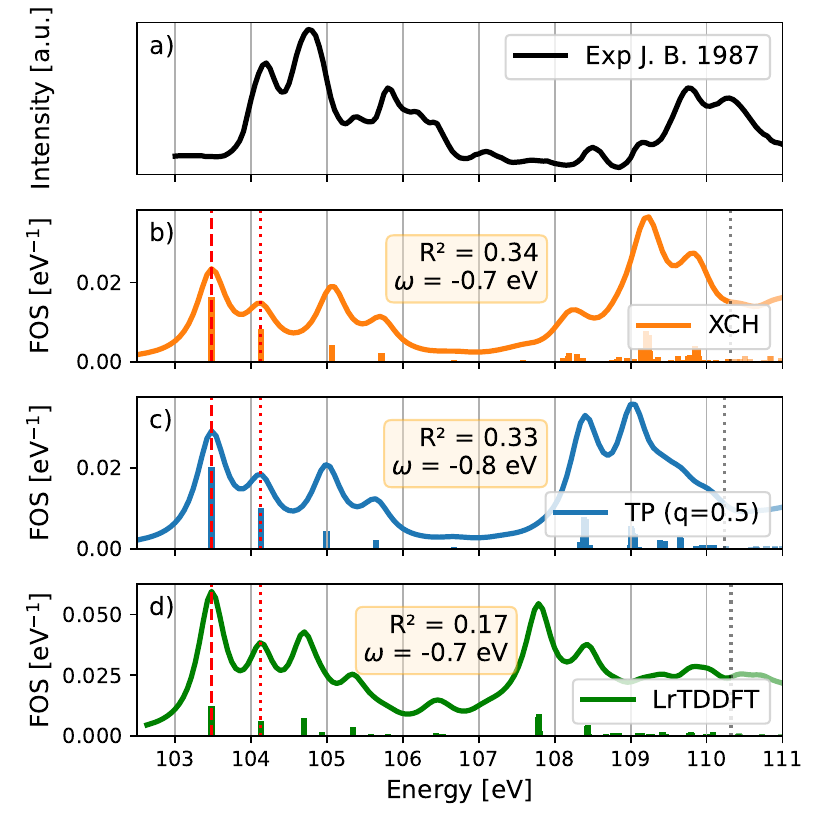}
    \caption{Si 2$p$ or $L_{2,3}$-edge NEXAFS spectra for SiCl$_4$. The experimental spectrum \changed{in a)} is adapted from Ref.~\onlinecite{BOZEK198733}.
    \changed{The methods used to calculate the spectra in b-d) are indicated in the legends.}
    }
    \label{fig:SiCl4}
\end{figure}
The Si $L_{2,3}$ edge spectra of SiCl$_4$ are shown in Fig.~\ref{fig:SiCl4}.  
Here, the experimental spectrum is very structured, with clearly observable peaks in the range
from 104 to 111~eV.
There are two regions with relatively high intensity separated by a region of
lower intensity within this energy range.
All simulations represent this general shape of the experimental data,
where the low-intensity region is the smallest in the LrTDDFT spectrum.
The $R^2$ values are relatively small due to the mismatch of the peak intensities,
where the simulations give a larger weight to the higher-energy peaks compared to experiment.
The shift $\omega$ that must be applied to the experimental spectrum
to maximize R$^2$ is very similar between -0.7 and -0.8 eV for all calculations.

\subsection{Solids}

We then consider solids, which we describe in a bulk approximation without the consideration of 
a surface. We solely use single-electron methods for solids due to the limitations of LrDDFT in the application to solids.\cite{Byun_2020,botti_long-range_2004}
We first examine the Ti $L_{2,3}$ edge or the Ti 2$p$ core-transitions. 
We analyze three different solid systems that feature a Ti 2$p$ core hole
in the following: SrTiO$_3$, the rutile structure of TiO$_2$, and the anatase structure of TiO$_2$.

\begin{figure}[!b]
    \includegraphics[width=0.5\textwidth]{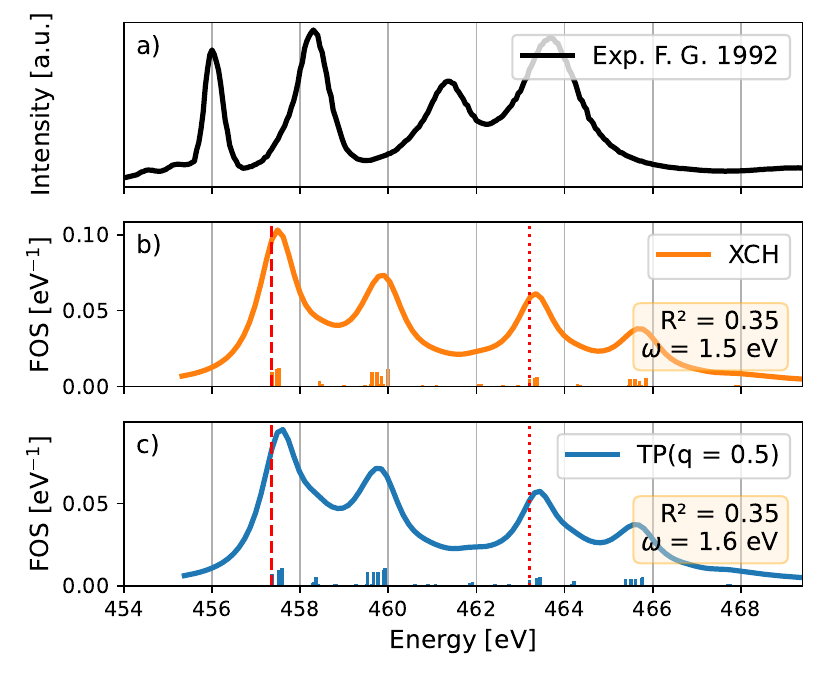}
    \caption{Ti 2$p$ or the $L_{2,3}$-edge NEXAFS spectra for solid SrTiO$_3$. The experimental spectrum \changed{in a)} is adapted from Ref.~\onlinecite{de_groot_2p_1992}.
    \changed{The methods used to calculate the spectra in b,c) are indicated in the legends.}
    }
    \label{fig: SrTiO3}
\end{figure}
\begin{figure*}[htb]
    \includegraphics[width=0.49\textwidth]{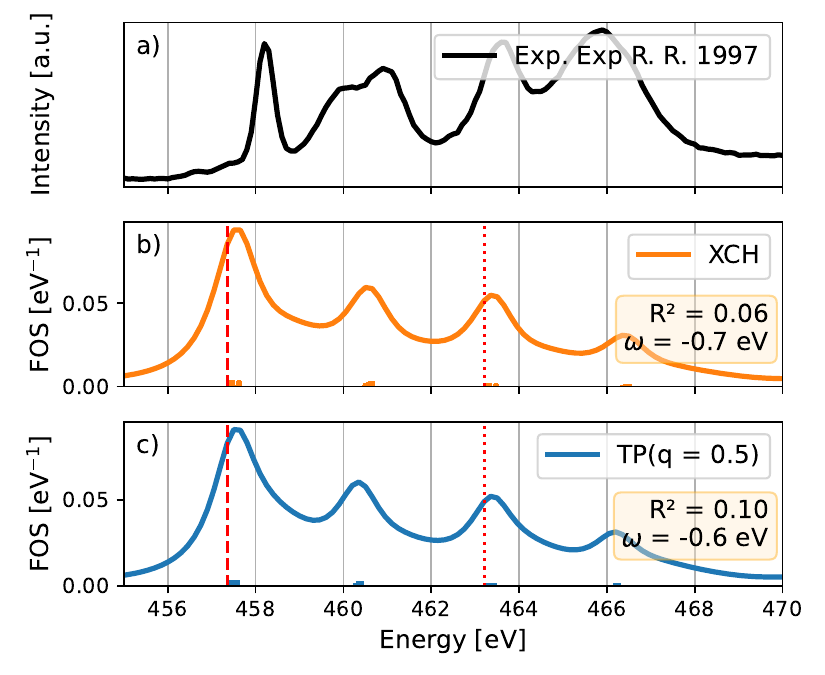}
    \includegraphics[width=0.49\textwidth]{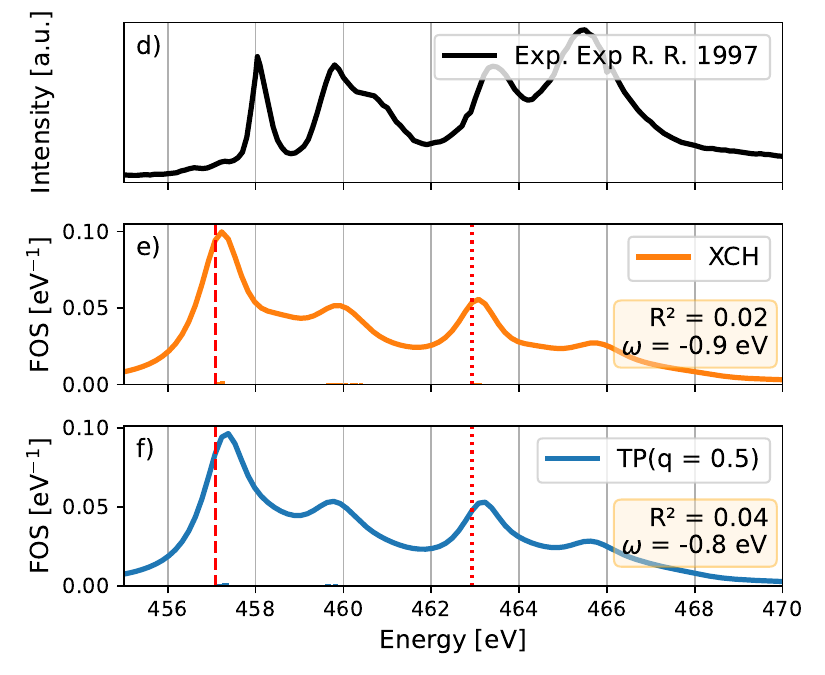}
    \caption{Ti 2$p$ or the $L_{2,3}$-edge NEXAFS spectra for solid TiO$_2$ a-c) rutile 
    and d-f) anatase structure. The experimental spectra \changed{in a,d)} are adapted from Ref.~\onlinecite{TiO21997}.
\changed{The methods used to calculate the spectra in b,c,e,f) are indicated in the legends.}
    }
    \label{fig: XAS TiO2 R}
\end{figure*}
The unit cell of SrTiO$_3$ has a $Pm\bar{3}m$ structure. 
The initial structure was obtained from the Materials Project (MP)\cite{Materials_prject_main} 
with ID: mp-5229 and 
subsequently re-relaxed in a $[3\times3\times3]$ supercell. 
The Brillouin zone was sampled using three $k$-points in each direction within the supercell.
The Ti 2$p$ XAS spectrum of SrTiO$_3$, has been investigated
experimentally\cite{de_groot_2p_1992} and theoretically\cite{SrTiO3_xas_2023,van_benthem_bulk_2001,van_benthem_core-hole_2003a,van_benthem_core-hole_2003b} in great detail in the literature.
The spectra are presented in Fig.~\ref{fig: SrTiO3}, where the experimental spectrum\cite{de_groot_2p_1992}
consists of four main peaks with varying widths. 
Both TP and XCH qualitatively reproduce this general peak structure.

A more detailed investigation of the first peak unveils the similarity of this peak with
the Ti 3$d$ contributions in the spectrum of TiCl$_4$ in Fig.~\ref{fig: TiCl4}.
The multiplet effect not covered by the single-particle approaches is present also in SrTiO$_3$. 
While the explicit core-hole methods XCH and TP are able to describe the overall 
structure of the four peaks successfully, the multiplet effect needs many-body approaches
such as configuration interaction\cite{ikeno_multiplet_2009} or solutions of the
Bethe-Salpeter equation\cite{SrTiO3_xas_2023}.
We note, however,
that a 20.4~eV shift had to be applied there to align the theoretical and experimental first peaks.

We next turn to TiO$_2$ with its most common rutile and anatase crystal structures. The TiO$_2$ rutile structure belongs to the $Pm_4/mnm$ space group, where the initial structure from MP with ID mp-2657 was re-relaxed in the $2\times2\times3$ supercell. The Brillouin zone was sampled by five $k$-points in each direction within the supercell. The TiO$_2$ anatase structure is part of the $I4_1/amd$ space group, where the initial structure from MP with ID mp-390 was re-relaxed in the $3\times3\times1$ supercell. The Brillouin zone was also sampled by five $k$-points in each direction within the supercell. 
The calculated spectra for rutile are shown in Fig.~\ref{fig: XAS TiO2 R} a) and b), while c) displays an experimental spectrum reported in Ref.~\onlinecite{TiO21997}. In Fig.~\ref{fig: XAS TiO2 R} d), e), and f), we present similar data for the anatase structure, using the experimental spectra from the same study. 

The R$^2$ values for these spectra are notably low, between $0.02 - 0.10$. However, we note that the four peaks are present in approximately the expected region. The theoretical spectra also indicate a shift towards lower energies. Since the core-hole methods successfully reproduce the four peaks, it can be argued that they somewhat replicate the spectra. Yet, we face similar issues as with SrTiO$_3$, as we did not account for many-body effects and charge-transfer multiplet effects. Additionally, the theoretical spectra for the two structures appear quite similar. It is widely recognized that the best way to differentiate between rutile and anatase structures in XAS spectra is to analyze the peaks in the 459--462~eV range; this is where their only differentiation lies. Anatase features a more pronounced first peak, whereas rutile displays a stronger second peak\cite{TiO21997}. This distinction is somewhat observable in that range, yet we cannot definitively separate these TiO$_2$ structures using core-hole methods.

\begin{figure}[t!]
    \centering
    \includegraphics[width=\linewidth]{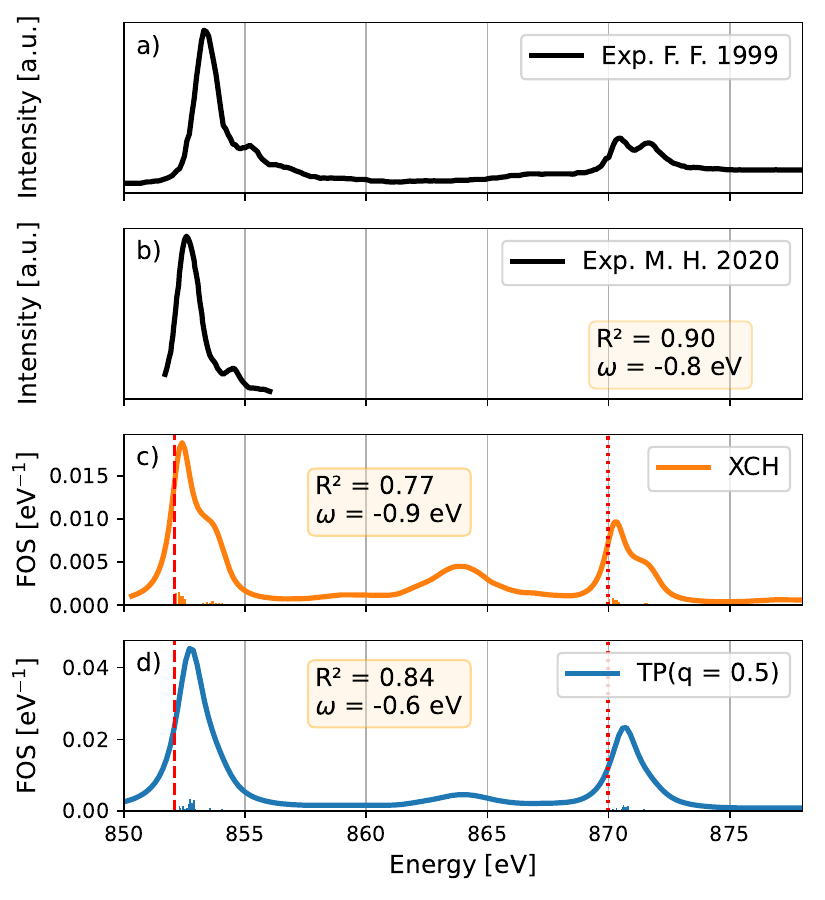}
    \caption{Ni 2$p$ or the $L_{2,3}$-edge NEXAFS spectra of solid NiO. The experimental spectrum in a) is adapted from Ref.~\onlinecite{Hepting2020}, and the experimental spectrum in b) is adapted from Ref.~\onlinecite{finazzi_2p3s3p_1999}.
    \changed{The methods used to calculate the spectra in c,d) are indicated in the legends.}
    }
    \label{fig: NiO XAS}
\end{figure}
We now consider NiO, an antiferromagnetic solid that exhibits a rocksalt crystal structure with a lattice constant of $a=4.19\text{\AA}$ \cite{Materials_prject_main, Moriyama2018}.
We use a $2\times2\times2$ cubic supercell and sample the Brillouin zone 
with five $k$-points in each direction within the supercell.

The spectra are displayed in Fig.~\ref{fig: NiO XAS}, where we display two spectra from the literature\cite{finazzi_2p3s3p_1999, Hepting2020} that encompass different energy ranges. 
The two experimental spectra align rather well, as indicated by the high $R^2$ value and the minor shift of $\omega=-0.8$ eV required to align the energy scales. Both experiments show a prominent peak around 853~eV that is accompanied by at least one smaller peak at slightly higher energy. The Ni $2p_{1/2}$ contribution is visible above 870 eV.

When comparing the two theoretical spectra to the experimental spectra in d), we notice that they are both in quite good agreement. They reproduce the general shape of the spectra; however, the proportions in the XCH in b) seem exaggerated, while the TP method offers a more accurate representation. This is also reflected in the R$^2$ values. The comparison between the theoretical spectra and the experimental results in c) can be found in Sec.~VI in SM.

However, Ikeno et al. also investigated the NiO spectra, as Ni is a transition metal, and found that by incorporating the CI method and multiplet effects, a better agreement with the experiments was achieved\cite{ikeno_multiplet_2009}. Although these methods produce improved spectra, the agreement derived from the TP is adequate to characterize the NiO structure; we believe this is due to the fact that, unlike the Ti structures, the NiO $d$-bands are partly filled, whereas the $d$-bands for the Ti structures are entirely open. 

\begin{figure}[!t]
\includegraphics[width=\linewidth]{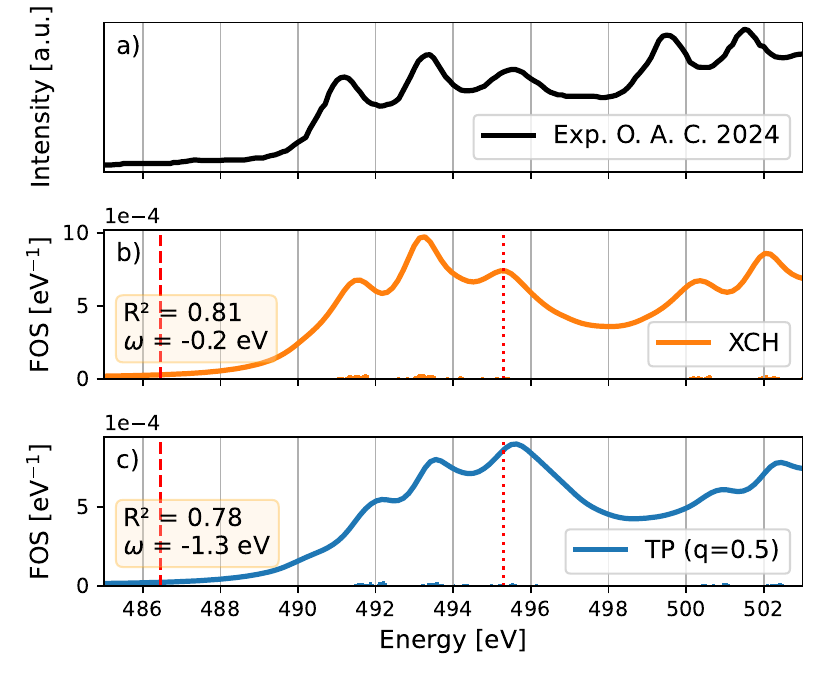} 
\caption{\label{fig:SnO2 T XAS} Sn 3$d$ or the $M_{3,4}$-edge NEXAFS spectra of SnO$_2$. 
The experimental spectrum \changed{in a)} is adapted from Ref.~\onlinecite{Sn_exp_XAS2024}.
\changed{The methods used to calculate the spectra in b,c) are indicated in the legends.}
}
\end{figure}

We also wish to address the $M_{4,5}$ NEXAFS spectra, specifically concerning $d$-type core holes. Here, we have selected tin as element and compared our calculations to the available experimental spectrum of stannic oxide (SnO$_2$). 
There are numerous gas-phase XPS spectra for Sn 3$d$, making the fit for the corresponding absolute shift $\delta$ 
unproblematic, as shown in Fig.~S3 in SM. Unfortunately, we have not found any molecular NEXAFS spectrum for comparison. Thus, we are limited to the bulk phase, excluding TDDFT.

SnO$_2$ exhibits a rutile crystal structure\cite{batzill_surface_2005} (space group $P$4$_2$/$mnm$) with lattice constants $a = b =$4.76~\AA, $c =$3.21~\AA. \cite{Materials_prject_main}.
The initial structure was sourced from the Materials Projects SnO$_2$, with the materials ID: mp-856.\cite{ Materials_prject_main}. Our calculation is done with a $2\times2\times3$ supercell with six $k$-points in each direction within the supercell; however, the calculations converge with four $k$-points. Further details can be found in Secs.~VII and VI in SM.

Fig.~\ref{fig:SnO2 T XAS} shows the spectra produced by the TP 
and XCH compared to the experimental spectra by Chuvenkova et al. \cite{Sn_exp_XAS2024}. 
Both theoretical spectra consistently reproduce the features of the experiment in a) with quite high R$^2$ values, with the XCH method demonstrating the best agreement.  The XCH spectrum also has the smallest energy shift compared to the experiment.

\subsection{Electron energy-loss in graphene}

\begin{figure*}[htb]
\includegraphics[width=\linewidth]{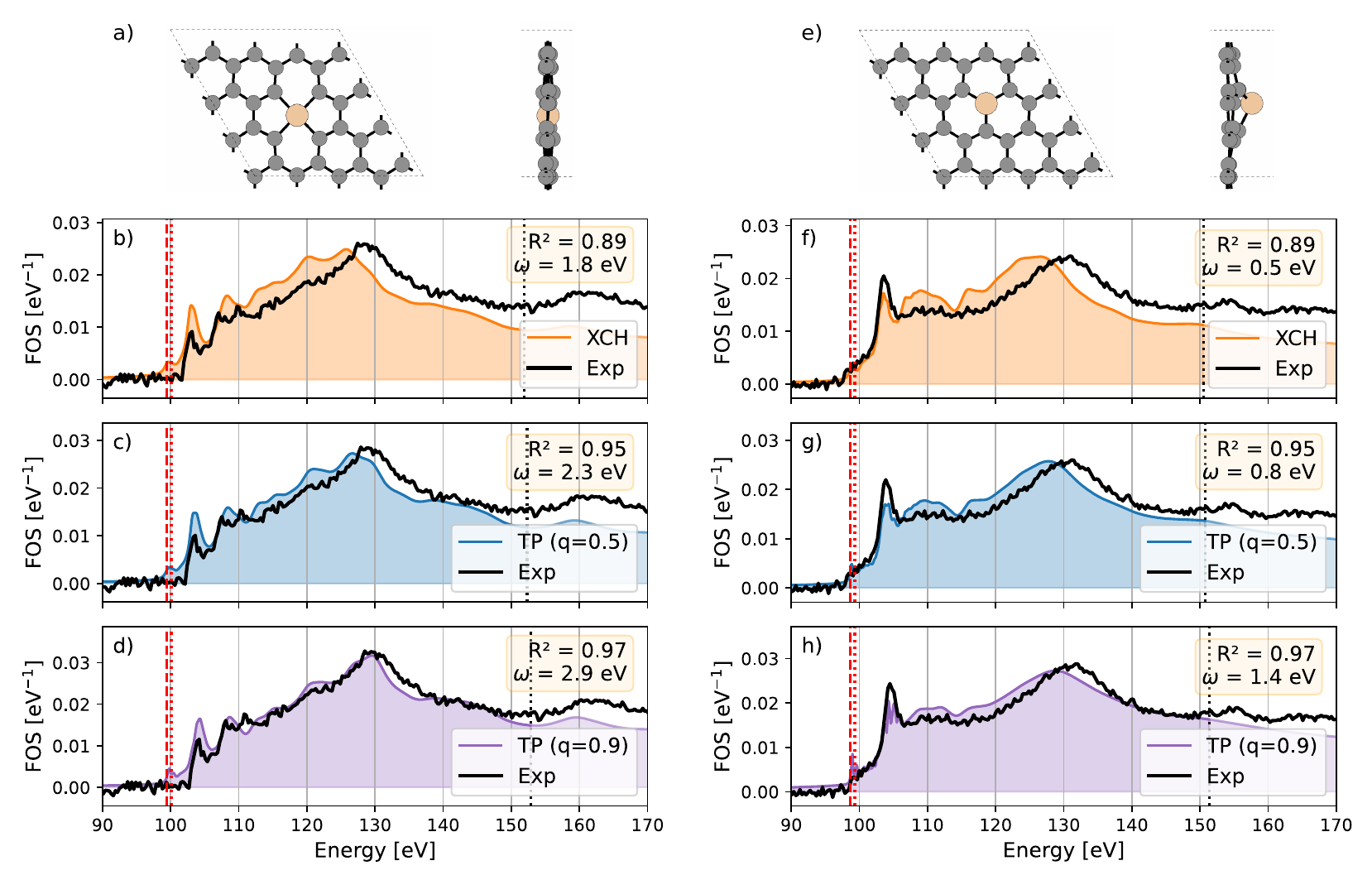} 
    \caption{(a and e) Structural model of a single Si atom covalently bound into a vacancy in single-layer graphene as a (a)~four-bonded and (e)~three-bonded substitution. 
    (b-d and f-h) Calculated Si 2$p$ or $L_{2,3}$-edge XAS spectra for a single Si atom in a graphene defect, represented in a) and e), compared to a single-atom EELS measurement. 
    The TP methods consider two core-hole occupancies of $q=0.5, 0.9$ \changed{as indicated in the legends in c,d,g,h).}
    The experimental spectra are from Ref.~\onlinecite{Ramasse_2013}.
    The experimental spectra are shifted
    by $\omega$ given in the caption to maximize R$^2$
    in each case.
}
\label{fig: ELLS Si}
\end{figure*}
We finally address the comparison of our calculations nominally meant for simulating XAS spectra to experimental electron energy-loss spectra of
single silicon atoms embedded in graphene defects.
Ramasse et al. have shown that these spectra can be used to distinguish
three- from four-bonded silicon atoms at defects\cite{Ramasse_2013}. 

The structures are modeled by a $4\times4$ single-layer graphene sheet with lattice constant of $a=2.46$ \AA \cite{Yang_2018}.
A  single Si atom substitutes a single C atom in the three-bonded configuration, while a single Si atom substitutes two C's in the four-bonded configuration. 
The Si atoms were displaced out of the plane and the structure was allowed to relax.
This lead to a back-relaxation into the plane for the four-bonded Si, while the three-bonded Si stayed out of the plane due to its larger size compared to the replaced C atom.
The relaxed structures are shown in Fig.~\ref{fig: ELLS Si} a) and e), respectively. 
The calculations are done in periodic boundary conditions.
The Brillouin zone is sampled by seven $k$-points in $x$ and $y$ directions while only using the Gamma point in $z$ direction. 

Fig. \ref{fig: ELLS Si} shows the spectra produced by the XCH method, the TP method with $q=0.5$ and the TP with $q=0.9$, for both the four- and three-bonded Si defect, in b-d) and f-h) respectively.
The agreement of the calculations with the experiment is of such quality, that the spectra can be 
directly overlaid, as reflected also by the large values of $R^2$.

Alignment of the first prominent peaks around 105~eV 
leads to a small mismatch of the broad peaks around 130~eV, where in particular the XCH is at slightly
lower energy.
This suggests and insufficient screening of the full core-hole in XCH compared to TP with $q=0.5$. Indeed, the further reduction of the core-hole charge
to $q=0.9$ helps in this respect, leading to 
very good agreement for both binding situations. 
The predicted spectra indeed allow to distinguish the binding mode,
as is explicitly shown in Fig.~S7 in SM.
Overall, it is remarkable how well the simulations are able to reproduce the EELS fine structure, as has also been observed previously without consideration of an explicit core-hole\cite{nicholls2013probing,Ramasse_2013,hardcastle2017robust,susi2017single-atom}. 

\section{Conclusions}

In conclusion, we have demonstrated a straightforward method to simulate
X-ray absorption spectra for core-states with finite orbital angular momentum.
The method is based on single-particle orbitals in the field of a core-hole
and does not require the explicit consideration of spin-orbit splitting.
We have compared the two mainly used approaches in the literature, excited core-hole (XCH) and transition-potential (TP),
that differ in the occupation of the core-hole state as well as the overall 
charge in the calculation.

Both the XCH and TP methods accurately replicate the $L_{2,3}$-edge XANES experimental spectra for the majority of the molecular systems examined.
Including the semi-empirical shifts, the spectra provided are on the absolute 
energy scale with deviations mostly below 1 eV. These asre in the range of possible 
experimental uncertainties as the comparison of experimental data for the same molecules shows.
The performance of TP and XCH is comparable to the LrTDDFT method and, in some instances, even surpasses it. 
In our comparison of the TP and XCH methods for molecular systems, we observed similar R$^2$ values, with the XCH method being slightly higher in Si cases and the TP method in S cases. For solid systems, TP slightly outperformed XCH in most instances, except for SnO$_2$, where XCH achieved better results.
The comparison with experimental EEL spectra is improved when the occupation of the core-hole is further reduced than in the traditional TP using a half occupied core-hole, however.

LrTDDFT showed a clear advantage over the single-particle methods with the Ti $L$-edge for TiCl$_4$, 
where the Ti has a fully open $d$-band due to the multiplet effect. 
This mixing of unoccupied states due to many-body effects is present only in LrTDDFT
and cannot be reproduced by single-particle methods.

A similar picture emerges for solid systems, where the core-hole methods also sufficiently reproduce the $L_{2,3}$ and $M_{4,5}$ edge XANES experimental spectra.
Also, at least for graphene, single-atom electron energy-loss spectra can be simulated with excellent accuracy.
However, we encountered similar challenges as with the molecular TiCl$_4$
due to missing multiplet effects that require a many-body description.

Overall, the explicit core-hole TP and XCH methods 
are able to sufficiently describe $L_{2,3}$-edge XANES for the majority of systems we evaluated.
There is no clear preference for either of the two methods as 
their advantages and disadvantages seem to be system-dependent.
Nevertheless, the methods described represent a computationally cheap way (a speed up factor of $\sim$ 40 relative to TDDFT) to
simulate XAS spectra with an overall good agreement to experiment 
on the absolute energy scale.
This may be very useful for a fast \changed{high-throughput} screening of spectra from a 
large group of materials as was applied e.g. for Raman spectra\cite{taghizadeh_library_2020}.

Our implementation has the advantage of being fully open
source, and we can thus expect such simulations to become
more widely accessible for the X-ray and electron microscopy
communities.

\section{Supplementary Material}

A comparison of the XPS and XAS nomenclature.
Details on the implementation, the determination of semiempirical shifts and the spin-orbit coupling calculations. List of the 
constants used. Details about the
optimal fitting of the spectra. Convergence tests.

\begin{acknowledgments}
E. J. and M. W. thank L. Mayrhofer and M. Moseler for useful discussion and for reading the manuscript.
E. J., N. H. and M. W.  acknowledge funding by the Deutsche Forschungsgemeinschaft (DFG, German Research Foundation) via the Excellence Cluster LivMatS under Germany’s Excellence Strategy—EXC-2193/1— 390951807.
E.J. and M.W. are thankful for the computing resources provided by the state of Baden-Württemberg through bwHPC and the German Research Foundation through grant number INST 40/575-1
FUGG (NEMO and JUSTUS2 clusters). We thank Quentin Ramasse for providing the raw data for the experimental EEL spectra considered in this study.
\end{acknowledgments}

\section{Author declarations}
\subsection{Conflict of Interest}
The authors have no conflicts to disclose.

\subsection{Author Contributions}

Esther A. B. Johnsen: Theoretical analysis (equal); Calculations (lead); Writing – review \& editing (equal);
Naoki Horiuchi: Calculations (equal);
Toma Susi:  Writing – review \& editing (equal);
Michael Walter: Calculations (equal); Theoretical analysis (lead); Conceptualization (lead); Resources (lead); Supervision (lead); Writing – review \& editing (equal).

\section*{Data Availability Statement}

Data is available on request from the authors.
The code developed to perform these calculations is available in the GPAW library.


%

\end{document}